\documentclass[aps,prl,preprint,tightenlines,superscriptaddress,showpacs,byrevtex]{revtex4} 
\usepackage{graphicx} 
\usepackage{epsfig}  
\usepackage[dvips]{color} 
\usepackage[dvips]{pstcol} 
\usepackage{dcolumn}  
\graphicspath{{ps}}

\definecolor{red}{rgb}{1,0,0}
\definecolor{purple}{rgb}{1,0,1}
\definecolor{green}{rgb}{0,1,0}
\definecolor{blue}{rgb}{0,0,1}
\definecolor{lblue}{rgb}{0,0.2,0.6}
\definecolor{yellow}{rgb}{0.1,0.1,0} 

\begin{document} 

\preprint{\vbox{ \hbox{   }
                 \hbox{BELLE-CONF-0608}
                 }}

\title{ \quad\\[0.5cm]  Measurement of the ratio 
${\cal B}$($D^0 \to \pi^+\pi^-\pi^0$)/
${\cal B}$($D^0 \to K^-\pi^+\pi^0$)} 

\affiliation{Budker Institute of Nuclear Physics, Novosibirsk}
\affiliation{Chiba University, Chiba}
\affiliation{Chonnam National University, Kwangju}
\affiliation{University of Cincinnati, Cincinnati, Ohio 45221}
\affiliation{University of Frankfurt, Frankfurt}
\affiliation{The Graduate University for Advanced Studies, Hayama} 
\affiliation{Gyeongsang National University, Chinju}
\affiliation{University of Hawaii, Honolulu, Hawaii 96822}
\affiliation{High Energy Accelerator Research Organization (KEK), Tsukuba}
\affiliation{Hiroshima Institute of Technology, Hiroshima}
\affiliation{University of Illinois at Urbana-Champaign, Urbana, Illinois 61801}
\affiliation{Institute of High Energy Physics, Chinese Academy of Sciences, Beijing}
\affiliation{Institute of High Energy Physics, Vienna}
\affiliation{Institute of High Energy Physics, Protvino}
\affiliation{Institute for Theoretical and Experimental Physics, Moscow}
\affiliation{J. Stefan Institute, Ljubljana}
\affiliation{Kanagawa University, Yokohama}
\affiliation{Korea University, Seoul}
\affiliation{Kyoto University, Kyoto}
\affiliation{Kyungpook National University, Taegu}
\affiliation{Swiss Federal Institute of Technology of Lausanne, EPFL, Lausanne}
\affiliation{University of Ljubljana, Ljubljana}
\affiliation{University of Maribor, Maribor}
\affiliation{University of Melbourne, Victoria}
\affiliation{Nagoya University, Nagoya}
\affiliation{Nara Women's University, Nara}
\affiliation{National Central University, Chung-li}
\affiliation{National United University, Miao Li}
\affiliation{Department of Physics, National Taiwan University, Taipei}
\affiliation{H. Niewodniczanski Institute of Nuclear Physics, Krakow}
\affiliation{Nippon Dental University, Niigata}
\affiliation{Niigata University, Niigata}
\affiliation{University of Nova Gorica, Nova Gorica}
\affiliation{Osaka City University, Osaka}
\affiliation{Osaka University, Osaka}
\affiliation{Panjab University, Chandigarh}
\affiliation{Peking University, Beijing}
\affiliation{University of Pittsburgh, Pittsburgh, Pennsylvania 15260}
\affiliation{Princeton University, Princeton, New Jersey 08544}
\affiliation{RIKEN BNL Research Center, Upton, New York 11973}
\affiliation{Saga University, Saga}
\affiliation{University of Science and Technology of China, Hefei}
\affiliation{Seoul National University, Seoul}
\affiliation{Shinshu University, Nagano}
\affiliation{Sungkyunkwan University, Suwon}
\affiliation{University of Sydney, Sydney NSW}
\affiliation{Tata Institute of Fundamental Research, Bombay}
\affiliation{Toho University, Funabashi}
\affiliation{Tohoku Gakuin University, Tagajo}
\affiliation{Tohoku University, Sendai}
\affiliation{Department of Physics, University of Tokyo, Tokyo}
\affiliation{Tokyo Institute of Technology, Tokyo}
\affiliation{Tokyo Metropolitan University, Tokyo}
\affiliation{Tokyo University of Agriculture and Technology, Tokyo}
\affiliation{Toyama National College of Maritime Technology, Toyama}
\affiliation{University of Tsukuba, Tsukuba}
\affiliation{Virginia Polytechnic Institute and State University, Blacksburg, Virginia 24061}
\affiliation{Yonsei University, Seoul}
  \author{K.~Abe}\affiliation{High Energy Accelerator Research Organization (KEK), Tsukuba} 
  \author{K.~Abe}\affiliation{Tohoku Gakuin University, Tagajo} 
  \author{I.~Adachi}\affiliation{High Energy Accelerator Research Organization (KEK), Tsukuba} 
  \author{H.~Aihara}\affiliation{Department of Physics, University of Tokyo, Tokyo} 
  \author{D.~Anipko}\affiliation{Budker Institute of Nuclear Physics, Novosibirsk} 
  \author{K.~Aoki}\affiliation{Nagoya University, Nagoya} 
  \author{T.~Arakawa}\affiliation{Niigata University, Niigata} 
  \author{K.~Arinstein}\affiliation{Budker Institute of Nuclear Physics, Novosibirsk} 
  \author{Y.~Asano}\affiliation{University of Tsukuba, Tsukuba} 
  \author{T.~Aso}\affiliation{Toyama National College of Maritime Technology, Toyama} 
% ...
\collaboration{The Belle Collaboration}

\collaboration{Belle Collaboration} 
\noaffiliation 

\begin{abstract} 
We report a high-statistics measurement of 
the relative branching fraction  
${\cal B}$($D^0 \to \pi^+\pi^-\pi^0$)/
${\cal B}$($D^0 \to K^-\pi^+\pi^0$).  
A 357 fb$^{-1}$ data sample collected with the Belle detector at the 
KEKB asymmetric-energy  
$e^+ e^-$ collider was used for the analysis. The relative branching 
fraction  
${\cal B}$($D^0 \to \pi^+\pi^-\pi^0$)/
${\cal B}$($D^0 \to K^-\pi^+\pi^0$) is determined  
with an accuracy comparable to the latest world average value.  
\end{abstract} 


\maketitle 

\tighten 

{\renewcommand{\thefootnote}{\fnsymbol{footnote}}}
\setcounter{footnote}{0} 

\section{Introduction} 


This measurement is the first step towards a high-statistics 
Dalitz-plot analysis of the $D^0 \to \pi^+\pi^-\pi^0$ decay. 
The latter could give insight into the controversy on 
the S-wave $\pi^+\pi^-$ contribution in these decays \cite{cleo2,focus}, 
as well as a sensitive study of the $CP$ violation in the neutral D meson 
system. Knowledge of ${\cal B}$($D^0 \to \rho \pi$)/${\cal B}$($D^0 \to K^* K$) 
(also based on the $D^0 \to \pi^+\pi^-\pi^0$ Dalitz analysis) could improve our 
understanding of the apparent discrepancy of the measured two-body branching 
fractions ($D^0 \to$ KK, $\pi\pi$) with the theoretical expectations \cite{bucce}. 
The accuracy of the value of 
${\cal B}$($D^0 \to \pi^+\pi^-\pi^0$) as reported in PDG04 \cite{pdg04} 
is poor. Using a large data sample of $D^0$ decays accumulated with the Belle 
detector, we provide a 
significantly improved measurement using the $D^0 \to K^-\pi^+\pi^0$ 
decay mode for normalization. Since both decay modes involve a neutral 
pion and the same number of 
charged tracks in the final state, several sources of the systematic 
uncertainties are avoided in a determination of the relative branching 
fraction. 
The obtained result 
can then be compared to recent measurements by the CLEO \cite{cleo3} 
and BABAR \cite{babar} collaborations. A detailed study of the 
$D^0 \to \pi^+\pi^-\pi^0$ decay as well as of other $D^0$ $CP$-symmetric 
final states, can 
be used to further improve statistics for the measurement of the 
angle $\phi_3$ ($\gamma$) of the CKM-matrix . \\ 


\section{Experiment} 

The Belle detector is a large-solid-angle magnetic spectrometer 
located at the KEKB $e^+ e^-$ storage rings, which collide 8.0 GeV 
electrons with 3.5 GeV positrons and produce 
$\Upsilon$(4S) at the energy of 10.58 GeV. Closest to 
the interaction point is a silicon vertex detector (SVD), 
surrounded by a 50-layer central drift 
chamber (CDC), an array of aerogel Cherenkov counters (ACC), 
a barrel-like arrangement of 
time-of-flight (TOF) scintillation counters, and an electromagnetic 
calorimeter (ECL) comprised of CsI (Tl) crystals. These subdetectors 
are located inside a superconducting solenoid coil 
that provides a 1.5 T magnetic field. An iron flux-return yoke located 
outside the coil is 
instrumented to detect $K^0_L$ mesons and identify muons. 
The detector is described in detail elsewhere \cite{det,svd}. 


\section{Data selection}


For this analysis, we used a data sample of 357 fb$^{-1}$ accumulated 
with the Belle detector. $D^0$ candidates are selected from 
$D^* \to D^0 \pi_{\rm slow}$ decays where the charge of the 
$\pi_{\rm slow}$ tags the $D^0$ flavour: ${D^*}^{\pm} \ \to \ D^0/\overline{D^0} 
\pi^{\pm}_{\rm slow}$. $D^*$'s originate mainly from continuum. Although we 
do not apply any topological cuts, the yield of $D^*$'s from $B$-meson decays 
is negligible; such events are rejected by other kinematical cuts such as the 
strong $p_{\rm cms}$($D^*$) requirement. 
$D^0$ mesons are reconstructed from combinations of two oppositely 
charged pions (or a charged pion and kaon in the case of $K^-\pi^+\pi^0$) 
and one neutral pion. The latter is reconstructed from two $\gamma$ candidates 
satisfying the $\pi^0$ mass requirement given below. \\ 

The following kinematic criteria are applied to 
the charged track candidates: 
the distance from the nominal interaction point to the point of closest 
approach of the track is required to be within 0.15 cm in the radial 
direction ($dr$) and 0.3 cm 
along the beam direction ($dz$). We also require the transverse 
momentum of the track $p_{\perp}$ $>$ 0.050 GeV/c to suppress 
beam background. Kaons and pions are separated by combining the responses of
the ACC and the TOF with the $dE/dx$ measurement from the CDC to
form a likelihood $\mathcal{L}(h)$ where $h$ is a pion or a kaon.
Charged particles are identified as pions or kaons using the likelihood ratio
$\mathcal{R}_{\rm PID}=\mathcal{L}(K)/(\mathcal{L}(K)+\mathcal{L}(\pi))$.
For charged pion identification, we require $\mathcal{R}_{\rm PID}<0.4$.
This requirement selects
pions with an efficiency of 93\% and misidentified kaons with an
efficiency of 9\%.
For the identification of charged kaons, the requirement is
$\mathcal{R}_{\rm PID}>0.6$; in this case, the efficiency for kaon identification 
is 86\% and the probability to misidentify a pion is 4\%. \\

We impose conditions on the energies of the photons 
constituting the $\pi^0$ 
candidate ($E_{\gamma}(\pi^0)$ $>$ 0.060 GeV), the two-photon invariant mass 
($0.124$ GeV/c$^2 \ < M(\gamma\gamma) \ < \ 0.148$ GeV/c$^2$) 
and the $\pi^0$'s momentum in the laboratory frame 
($p_{\rm lab}$($\pi^0$)  $>$ 0.3 GeV/c)  
to suppress combinatorial $\pi^0$'s. \\ 

The mass difference of $D^*$ and $D^0$ candidates should satisfy the 
restrictions: 
$ 0.1442$ GeV/c$^2 \ < \ M(\pi_{\rm slow}\pi^+\pi^-\pi^0) - 
M(\pi^+\pi^-\pi^0) \ < \ 0.1468$ GeV/c$^2$ 
and $ 0.1440$ GeV/c$^2 \ < \ M(\pi_{\rm slow}K^-\pi^+\pi^0) 
- M(K^-\pi^+\pi^0) \ < \ 0.1470$ GeV/c$^2$. 
The momentum of the $D^*$ in the center-of-mass frame of the 
$\Upsilon(4{\rm S})$ 
must lie in the range: 
$ 3.5$ GeV/c$ \ < \ p_{\rm cms}(D^*) \ < \ 4.3$ GeV/c. 
The lower cut is applied to suppress slow fake $D^*$'s reconstructed 
from combinatorial background that originates from $B$ decays. 
The upper cut restricts $p_{\rm cms}(D^*)$ to the region 
where the Monte Carlo (MC) distribution is in good agreement with the data 
(see Fig. 1, left). 
To eliminate background from the 
$D^0 \to K_{\rm S}\pi^0 \to (\pi^+\pi^-)\pi^0$ decays, 
the following veto on $M^2(\pi^+\pi^-)$ is applied: 
$0.21$ GeV$^2$/c$^4 \ < \ M^2(\pi^+\pi^-) \ 
< \ 0.29$ GeV$^2$/c$^4$. \\ 

\section{Efficiency calculation} 

To obtain detection efficiencies, about 1.2$\times 10^6$ phase-space 
distributed MC events have been generated for each of the two modes, 
processed using the GEANT--based detector simulation \cite{geant} 
and reconstructed with the same selection criteria as the data. \\ 

To obtain a more realistic $D^0$ decay model than a 
uniform Dalitz-plot 
distribution, the events of both signal MC samples have been 
weighted using matrix elements based on decay models obtained by 
CLEO~\cite{cleo2} 
in the framework of a 3-resonance model 
($\rho^+$, $\rho^-$, $\rho^0$ and a nonresonant 
contribution) for $D^0 \to \pi^+\pi^-\pi^0$ and in the framework of a 
7-resonance model 
($\rho^+$, ${K^*}^-$, ${\bar{K^0}}^*$, $K_0(1430)^-$, ${\bar K}_0(1430)^0$, 
$\rho(1700)^+$, ${K^*}^-(1680)$ and a nonresonant contribution) for the case 
of $D^0 \to K^-\pi^+\pi^0$ \cite{kpp2} (see Fig. 2). \\ 

\begin{figure}
 \epsfig{figure=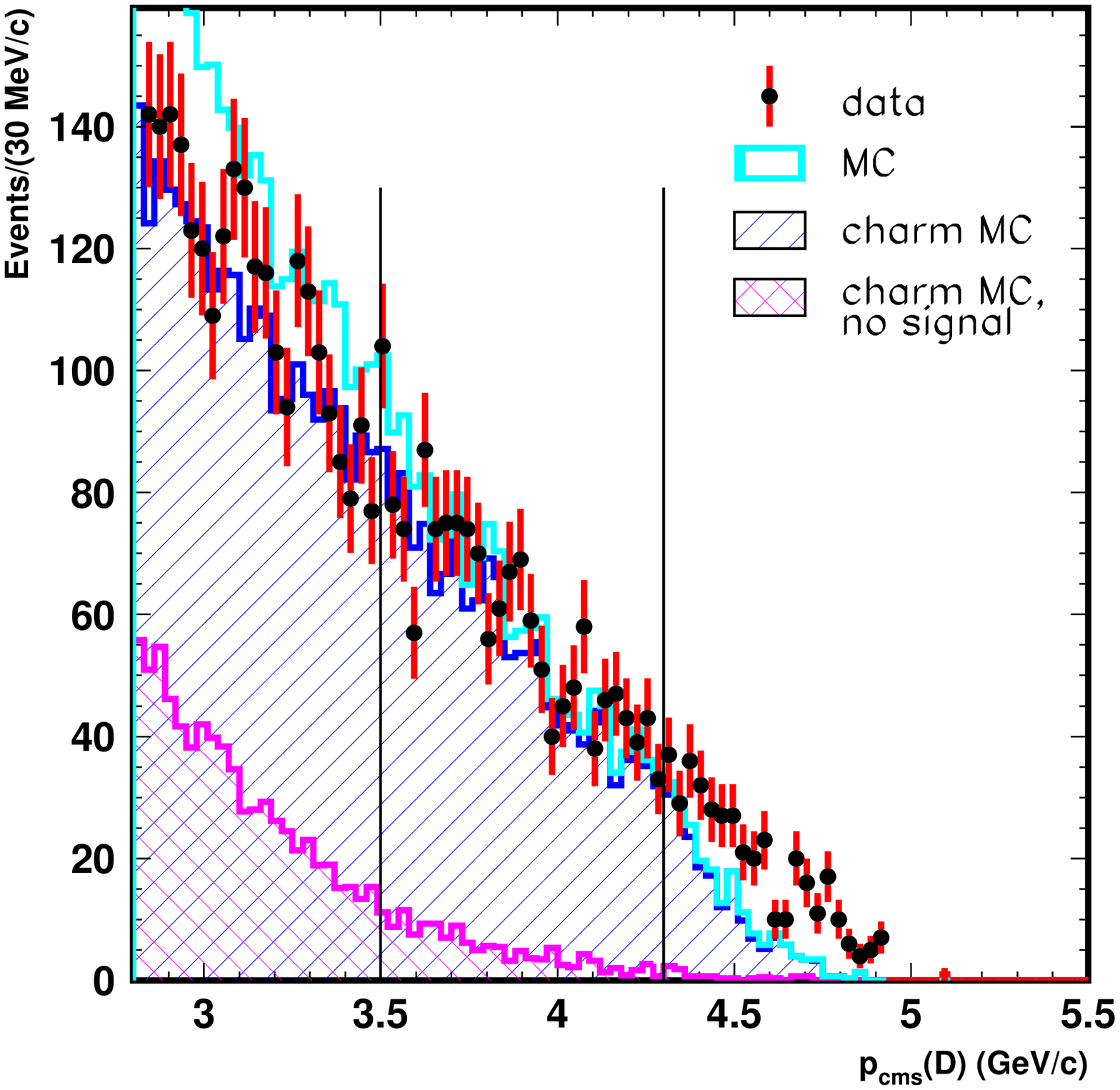,width=0.45\textwidth}  
 \epsfig{figure=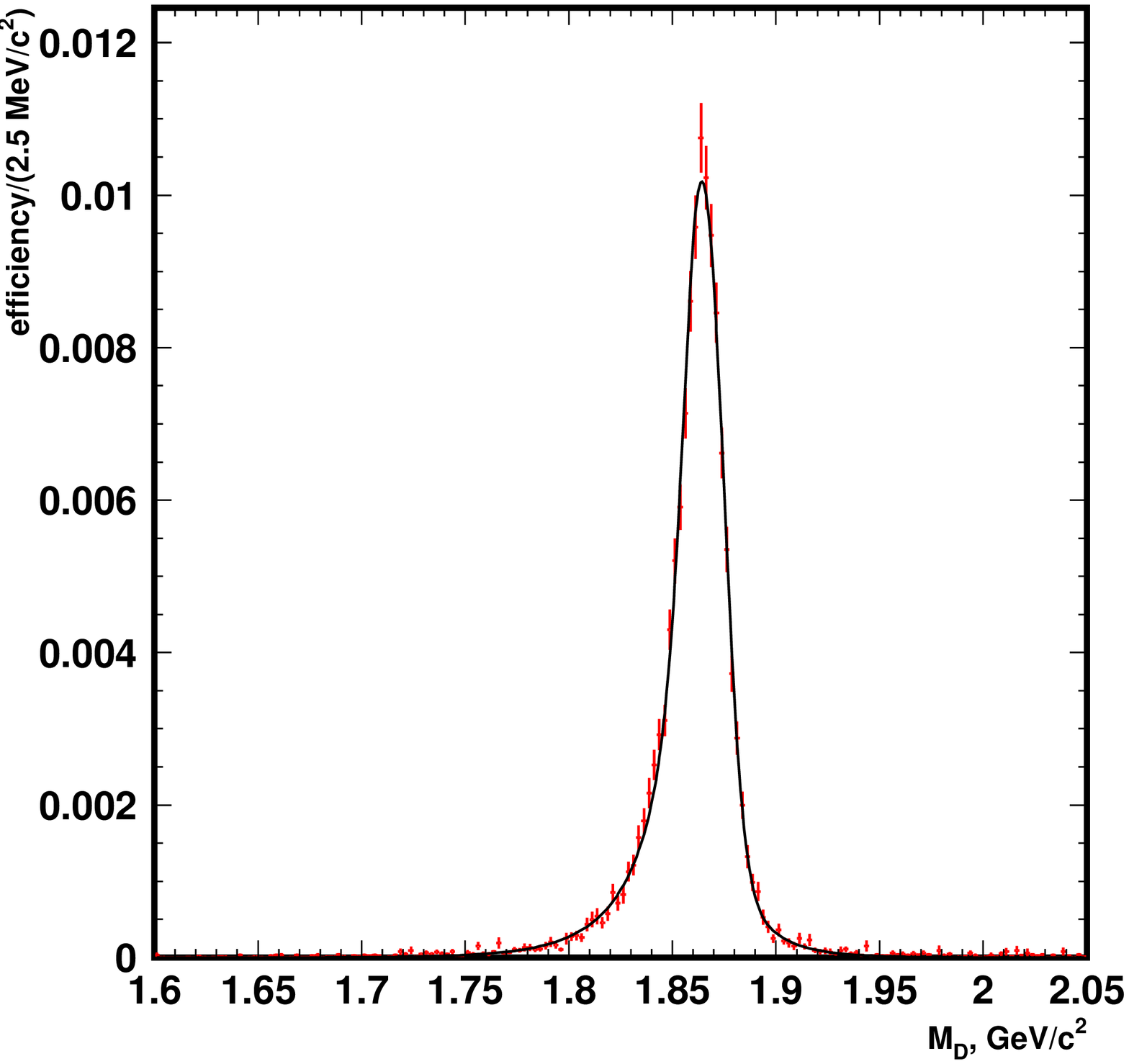,width=0.49\textwidth}  
 \caption{Left: $p_{\rm cms}$($D^*$) distribution  
 for experimental data (points with error bars), generic MC (solid line), 
charm (originating from $e^+e^- \to c\bar{c}$) MC including 
signal events    
(singly hatched histogram) and background charm MC (doubly hatched histogram).
  Right: $M(\pi^+\pi^-\pi^0)$ distribution, MC,  
 fitted with a bifurcated  hyperbolic Gaussian and a regular Gaussian.} 
\end{figure} 

\begin{figure}   
 \epsfig{figure=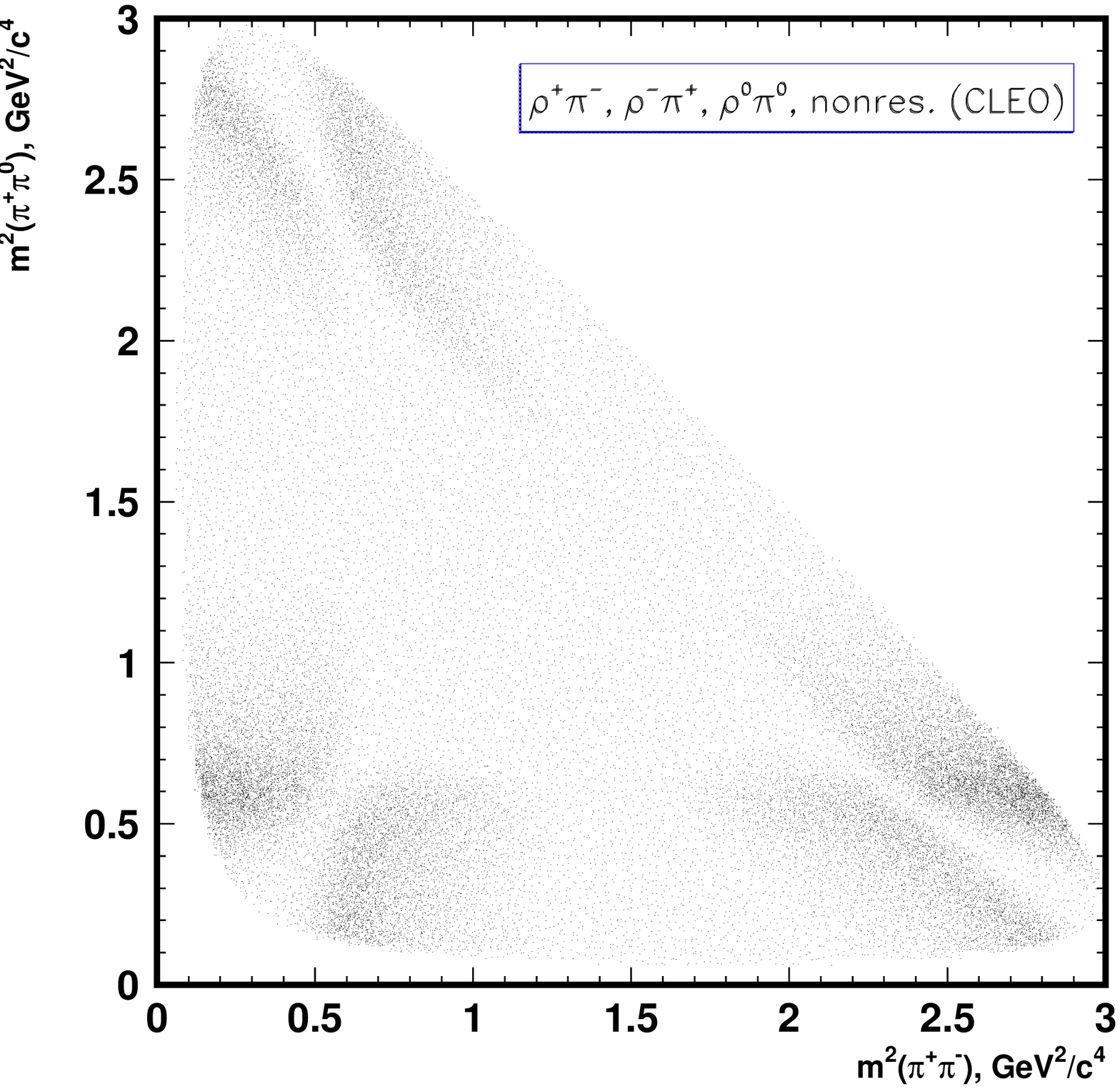,width=0.47\textwidth}  
 \epsfig{figure=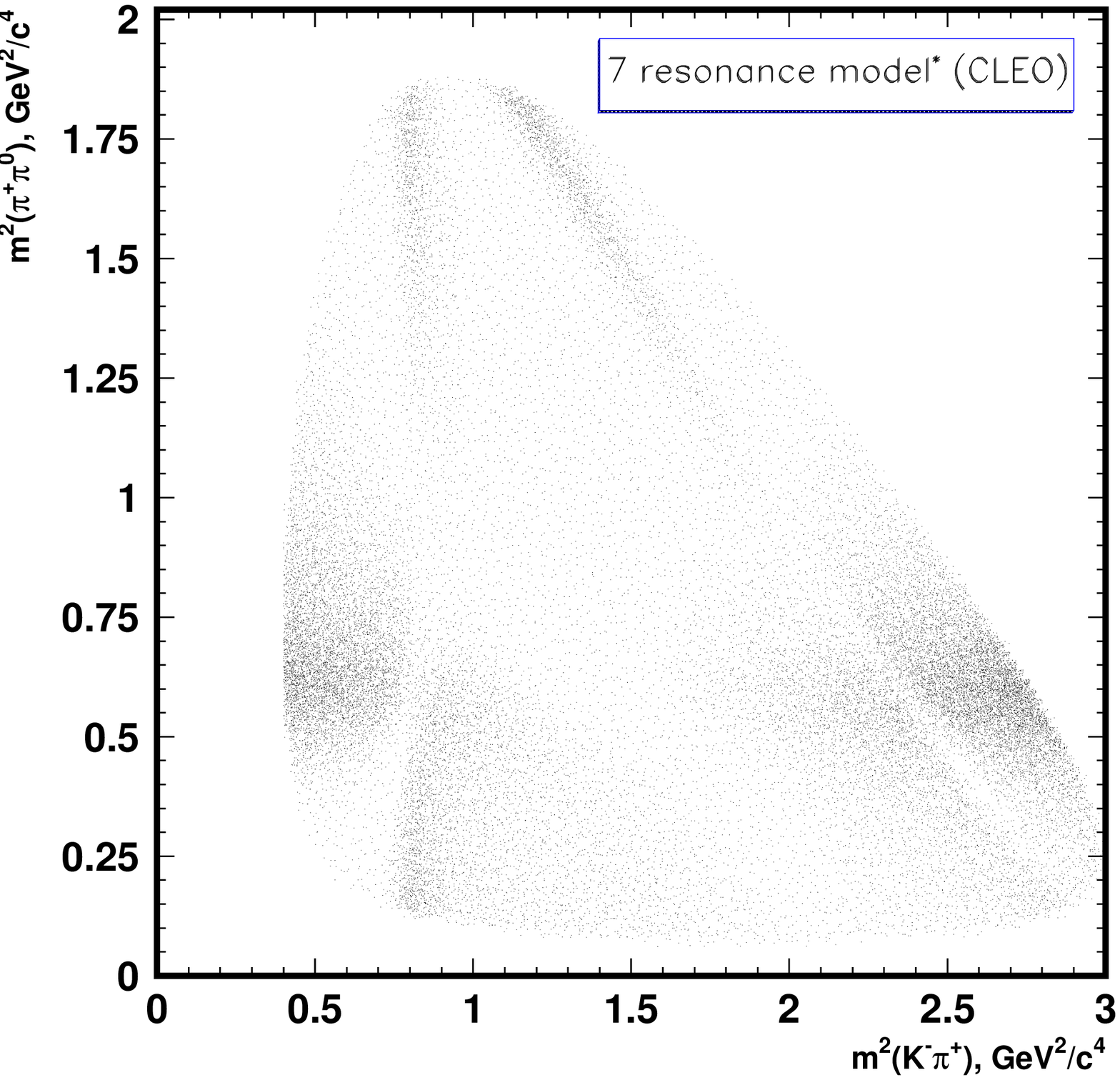,width=0.47\textwidth} 
 \caption{Dalitz plot distribution for "initial" signal MC 
(before skimming and detector 
 simulation) in the $D^0\to \pi^+\pi^-\pi^0$ (left, 3-resonance model) 
and $D^0\to 
 K^-\pi^+\pi^0$ 
 (right, 7-resonance model) cases.} 
\end{figure}

%

 The obtained yield is normalized to the same MC data but 
before detector simulation and before application of selection criteria. 
Since for the relative branching fraction 
measurement only the ratio of 
$\pi^+\pi^-\pi^0$ and $K^-\pi^+\pi^0$ efficiencies is needed, 
we only generate events in the 
$p_{\rm cms}({\rm D}^*)$ range from 3.5 to 4.3 GeV/c 
where MC is in good agreement with the data (see Fig. 1, left). 
The obtained efficiencies are: 
$\varepsilon(\pi^+\pi^-\pi^0) \ = \ (13.433 \pm 0.077)\%,$ 
$\varepsilon(K^-\pi^+\pi^0) \ = \ (13.065 \pm 0.074)\%.$ 

\section{Background description} 

To describe the background shape in the $M(\pi^+\pi^-\pi^0)$ 
signal region, 
a sample of generic MC, equivalent to $\sim$250 fb$^{-1}$, has been 
processed with the same selection criteria as data. The contributions 
of $u\bar{u}$, 
$d\bar{d}$, $s\bar{s}$, $c\bar{c}$ and $B\bar{B}$ backgrounds 
have been summed up. For the $c\bar{c}$ sample signal events were excluded. 
Figure 3 shows the individual contributions to the $M(\pi^+\pi^-\pi^0)$ 
distribution where also a partial data sample, corresponding to the 
luminosity of MC, is shown. 
The $M(\pi^+\pi^-\pi^0)$ background distribution is fitted 
with a $2^{nd}$ order polynomial multiplied by an error function 
(kaon misidentification region) plus a $1^{st}$ order polynomial 
(combinatoric background) 
and a small Gaussian peak in the signal region (see Fig. 3, right). 
The Gaussian contribution is mainly due to combinations of a correctly 
reconstructed $D^0$ and a random $\pi_{\rm slow}$ candidate. \\ 

\begin{figure}                                                                                    
 \epsfig{figure=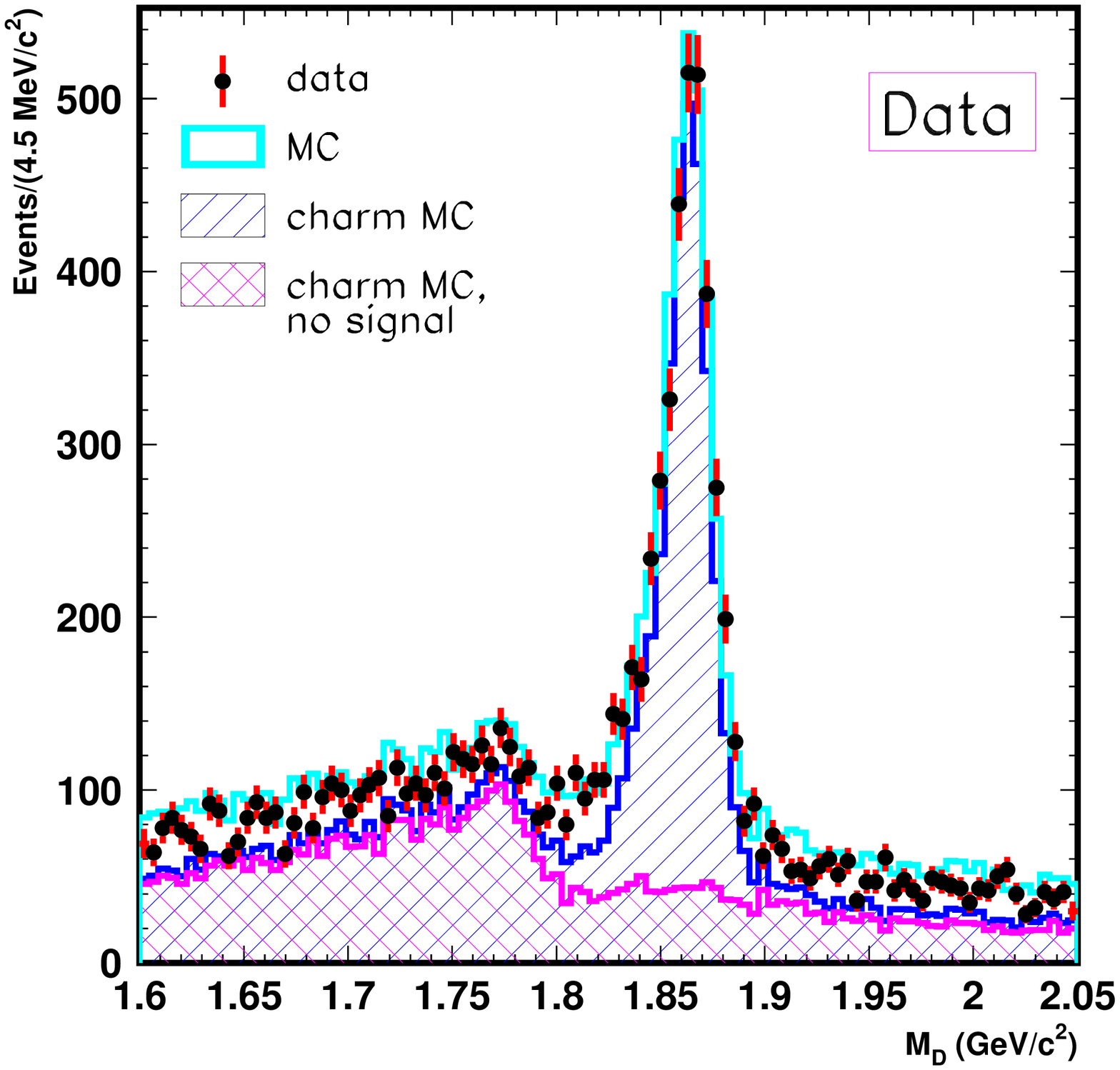,width=0.45\textwidth}   
 \epsfig{figure=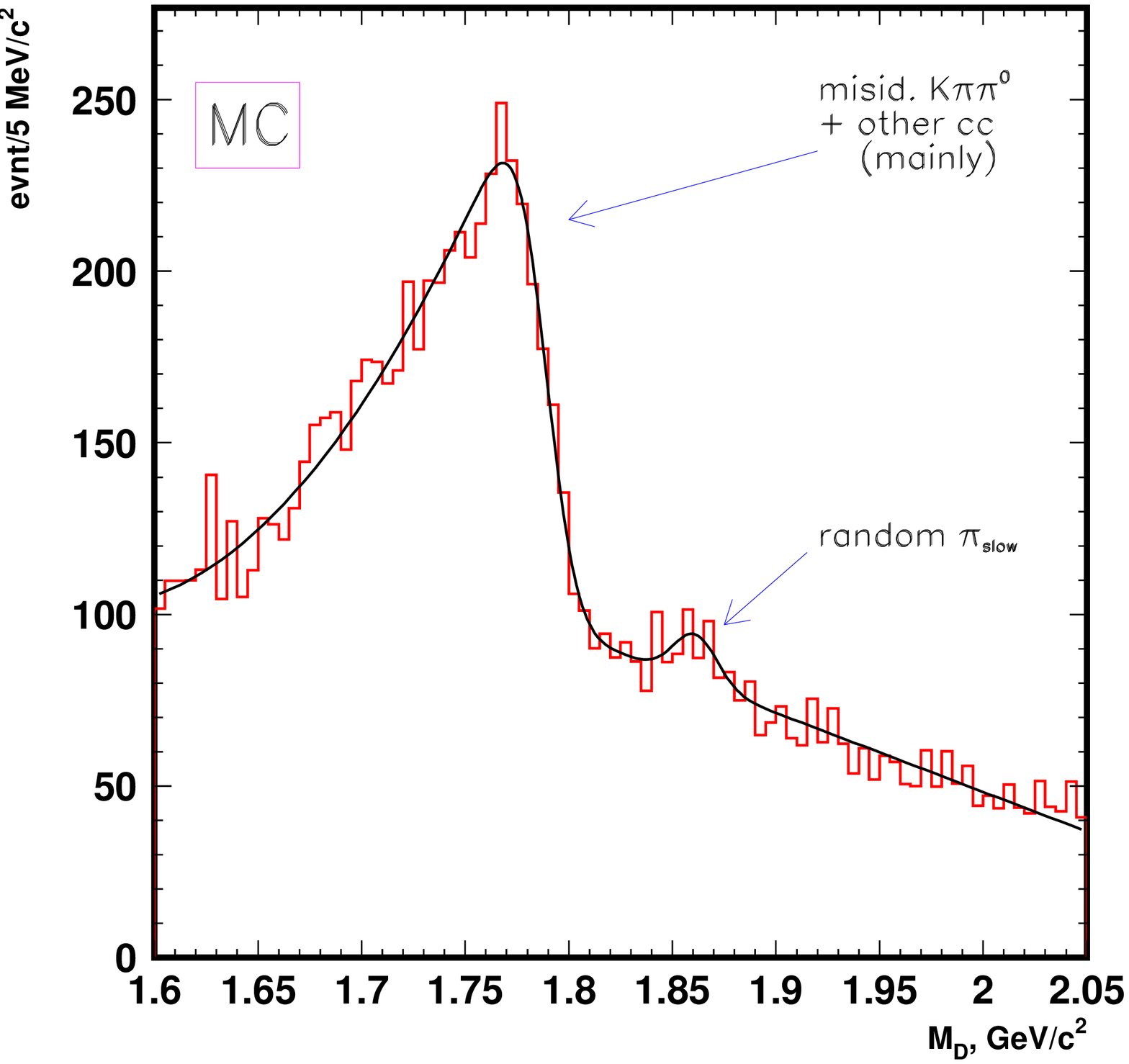,width=0.45\textwidth} 
 \caption{Left: $M(\pi^+\pi^-\pi^0)$ distribution  
 for experimental data (points with error bars), generic MC (solid line), charm    
 (originating from $e^+e^- \to c\bar{c}$) MC including signal events    
 (singly hatched histogram) and background charm MC (doubly hatched histogram).  
 Right: $M(\pi^+\pi^-\pi^0)$, "charm" MC with the signal excluded, 
 fitted to the function described in the text.} 
\end{figure}

 Most of the background is from e$^+$e$^- \to$ $c\bar{c}$; 
the $B\bar{B}$ background is negligible and 
$u\bar{u}$, $d\bar{d}$, $s\bar{s}$ backgrounds are linearly distributed 
in the $M(\pi^+\pi^-\pi^0)$ signal region. Among the $c\bar{c}$ background 
sources $D^* \to D^0(K\pi\pi^0)\pi$ is dominant: charged kaons are 
misidentified as pions and $M(\pi_{\rm fake}\pi\pi^0)$ is typically shifted downwards 
by $\sim 0.1$ GeV/c$^2$ thus being well separated from the signal. \\ 

 The background shape in the $M(K^-\pi^+\pi^0)$ 
distribution is obtained using a generic MC sample. 
The level of background in this mode is low but still has a nontrivial structure 
(see Fig. 4, right). The distribution has three distinctive peaks: the rightmost one is 
due to misidentified pions from $D^0 \to \pi^+\pi^-\pi^0$, the central one has the same 
origin as the one in the $D^0\to \pi^+\pi^-\pi^0$ case (random $\pi_{\rm slow}$), 
the leftmost feature originates mainly from $D^0 \to K n\pi$ where $n \ge 3$. 
The distribution is fitted by the sum of a $2^{nd}$ order polynomial and a $2^{nd}$ 
order polynomial multiplied by an error function (left peak) and two Gaussians. \\ 

\section{Signal fit} 

The shape of the signal peak in the experimental $\pi^+\pi^-\pi^0$ invariant mass distribution 
(see Fig. 5, left) is partially fixed to the MC one \cite{par}: the latter was fitted with 
a bifurcated hyperbolic Gaussian \cite{hyp, bif} and a regular Gaussian (see Fig. 1, right). 
The obtained shape with floating normalization and $\sigma$'s, together with 
the background shape with its floating normalization (i.e. 5 free 
parameters) is then used as the fit function for the 
measured $M(\pi^+\pi^-\pi^0)$ distribution. \\ 

\begin{figure}                                                                              
 \epsfig{figure=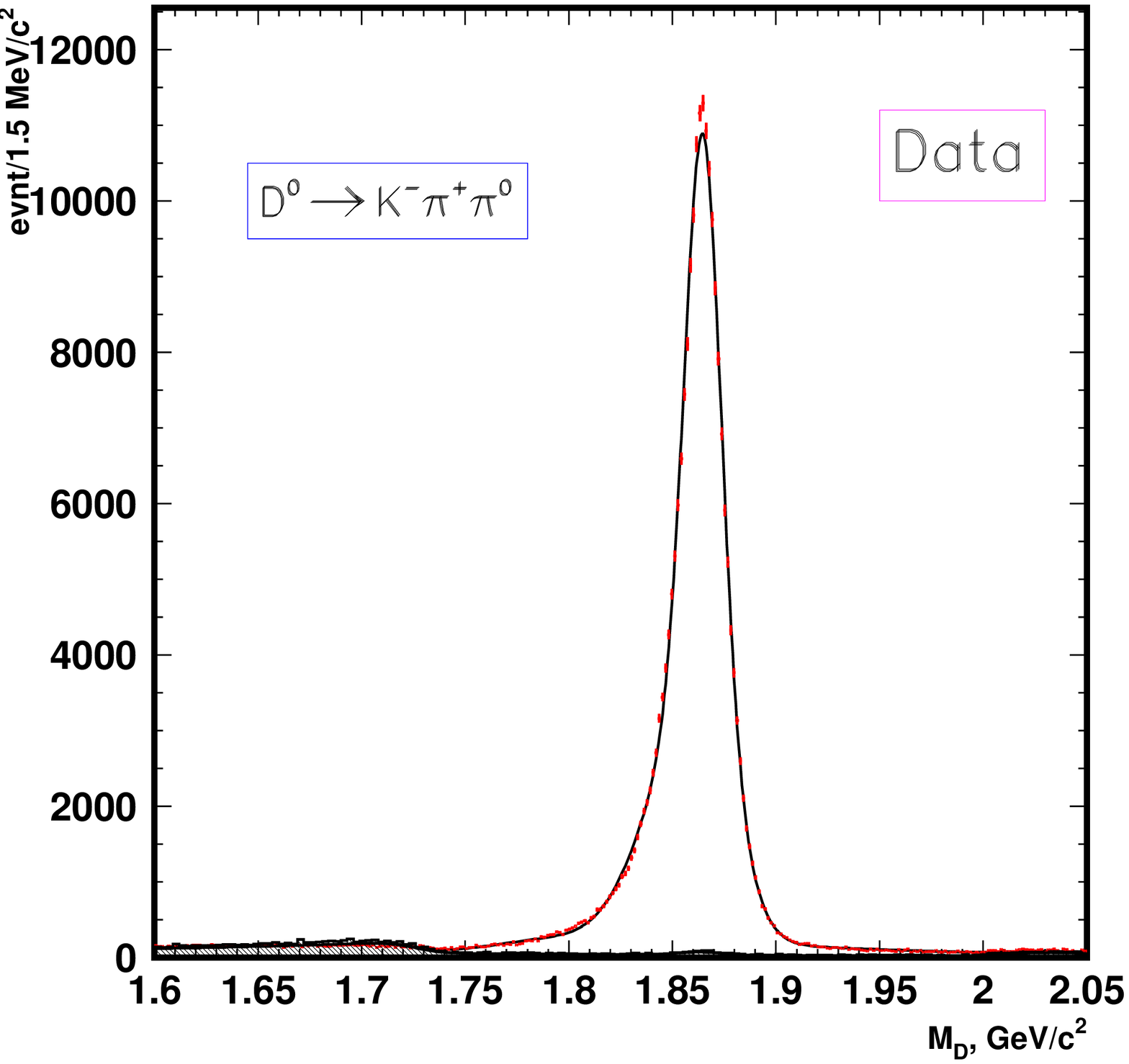,width=0.47\textwidth}                                  
 \epsfig{figure=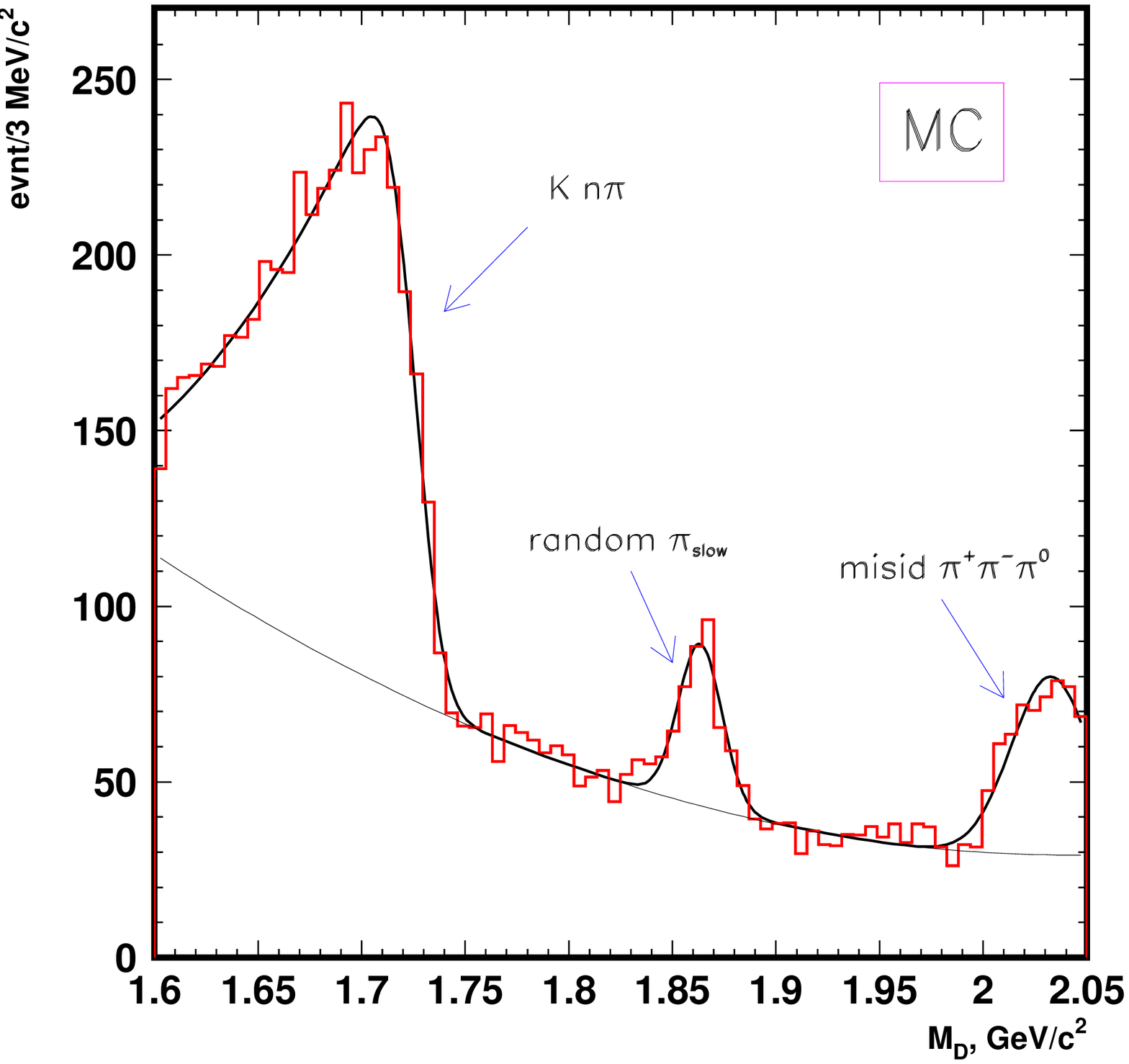,width=0.47\textwidth}                                
 \caption{Left: signal $M(K\pi\pi^0)$ distribution fitted with 2 bifurcated    
Gaussians + Gaussian (signal peak) and the generic MC shape (background). Right:   
background shape $M(K\pi\pi^0)$, generic MC. }  
\end{figure}

 To fit the experimental $K\pi\pi^0$ invariant mass distribution we use a $K\pi\pi^0$ 
background shape with floating normalization and the sum of two bifurcated and a regular 
Gaussian for the signal peak (see Fig. 4, right). 
Here, the parameters of the signal peak are free in the fit since the level of 
background is low and statistics are large enough. \\ 

\begin{figure}                                                                             
 \epsfig{figure=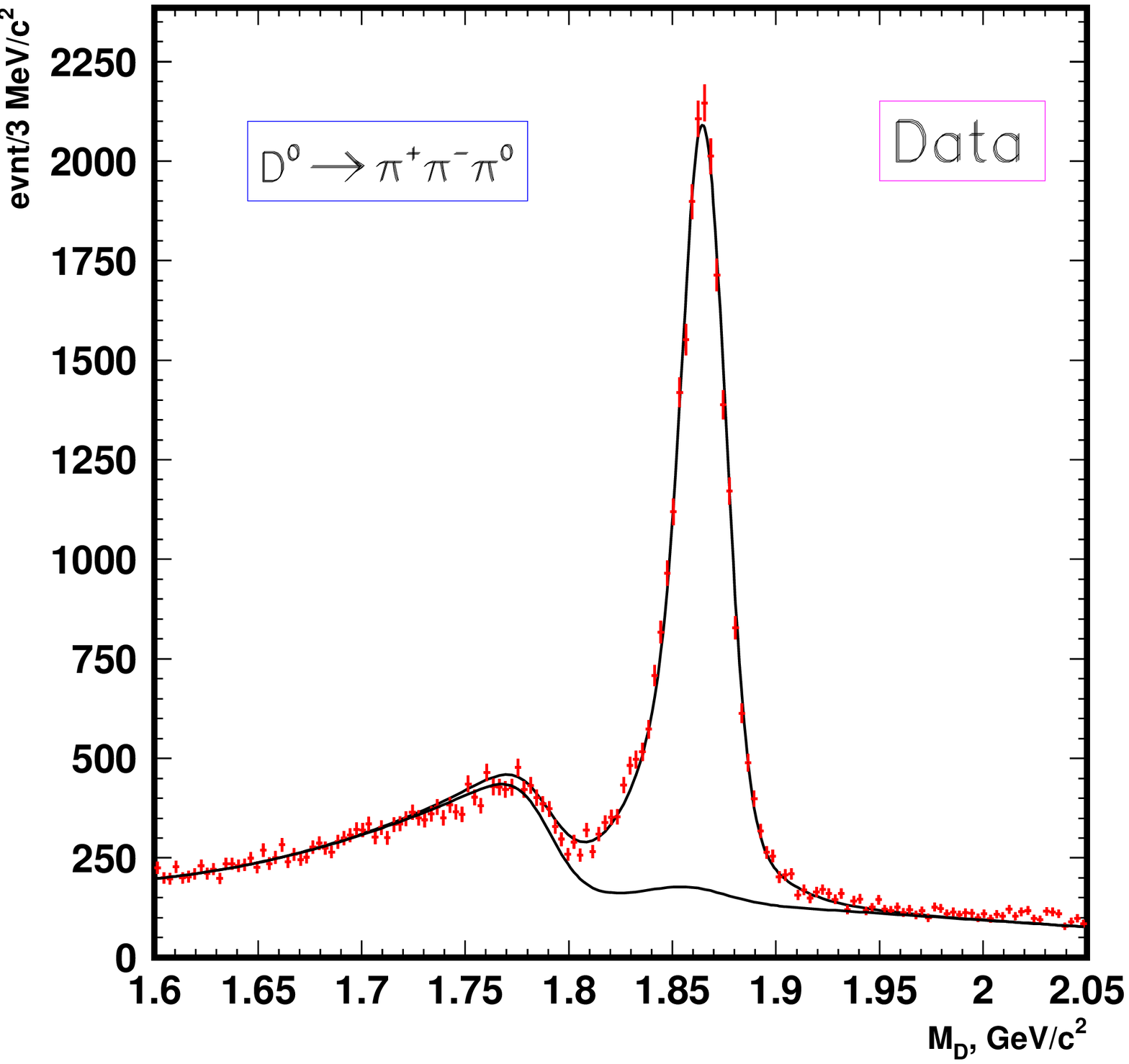,width=0.47\textwidth}                                 
 \epsfig{figure=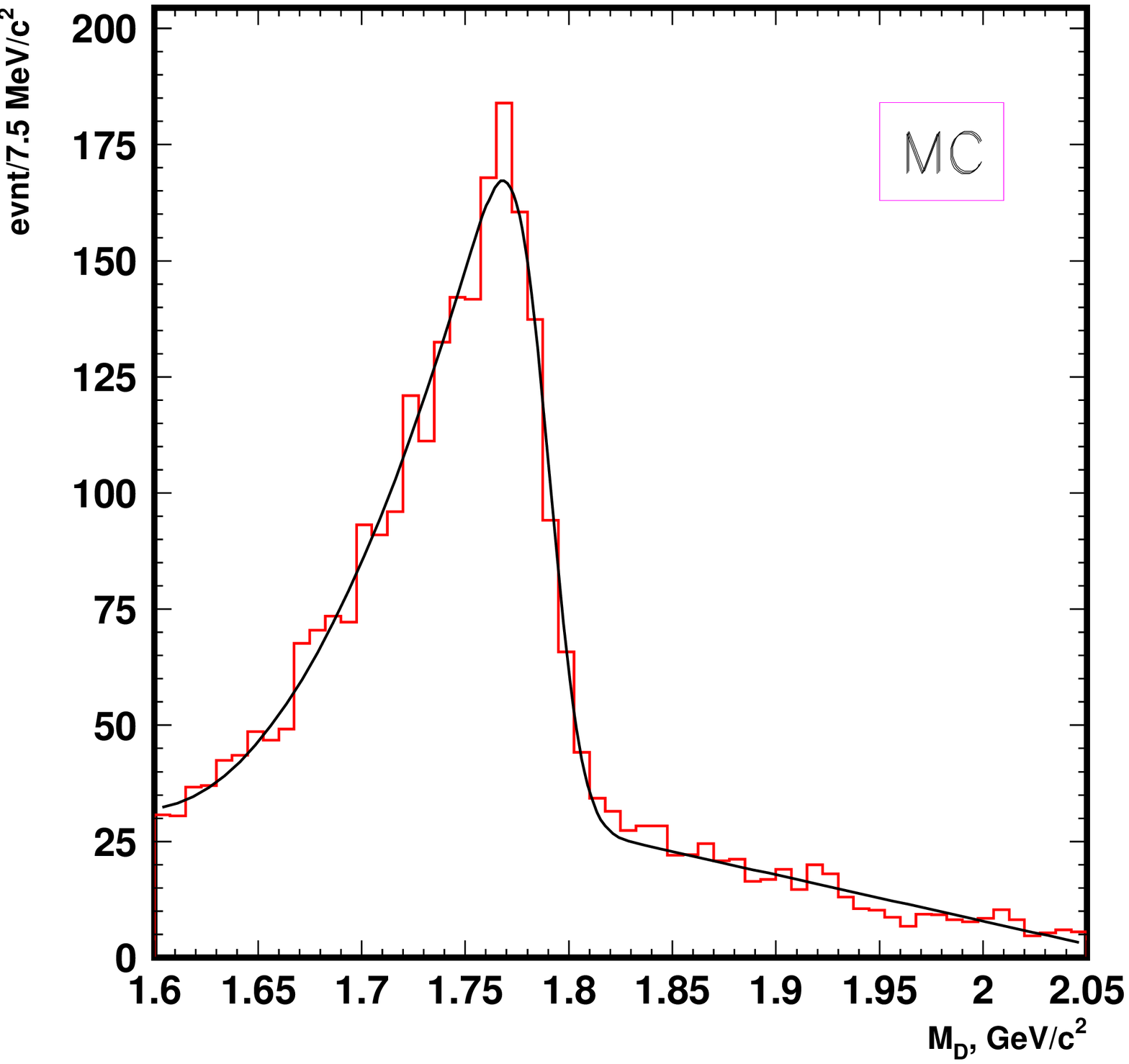,width=0.47\textwidth}                               
 \caption{Left: signal $M(\pi^+\pi^-\pi^0)$ distribution, fitted 
 to the signal MC shape (Fig. 1, right) for the signal peak and with the generic MC shape 
 (background). Right: misidentification shape, generic MC, used for an alternative 
 background fit to estimate the corresponding systematic error.}   
\end{figure}

 The yields of signal events in each channel, as obtained from the fit, 
are given in Table \ref{t7}: 

\begin{table}[ht!]                                                                               
\caption{Number of signal events and efficiencies} 
\vspace{0.5cm}                                                                                   
\label{t7} 
\begin{tabular}{|c|c|c|}   
 \hline  
 $D^0$ decay mode  & N$_{\rm ev}$ & $\varepsilon$, \% \\ \hline    
 $\pi^+\pi^-\pi^0$ & 22803 $\pm$ 203 & $13.433 \pm 0.077$ \\  
 $K^-\pi^+\pi^0$   & 237520 $\pm$ 500 & $13.065 \pm 0.074$ \\  \hline  
\end{tabular} \\
\end{table}


\section{Systematic uncertainties} 


Uncertainty on the tracking efficiency for the two charged tracks -   
$\pi^+\pi^-$ or $K^-\pi^+$ - 
cancels in the ratio of $D^0 \to \pi^+\pi^-\pi^0$ and $D^0 \to K^-\pi^+\pi^0$ 
branching fractions. The same holds for $\pi^0$ and slow pion (from $D^*$) 
reconstruction efficiencies. 
A possible difference in the efficiency of particle identification 
selection criteria 
between MC and data is taken into account as a correction to the 
branching ratio 
($r_K$/$r_{\pi} = 0.962 \pm 0.015$) and 
the uncertainty of this correction contributes to the systematic 
uncertainty of the result. 
The polar angle and momentum dependent data/MC corrections are measured 
independently using a large sample of $D^0 \to$ $K^-\pi^+$ decays. 
The uncertainty in the correction yields a systematic error of 1.6\%. \\ 

To estimate the systematic error due to the model used to describe the 
$D^0 \to \pi^+\pi^-\pi^0$ 
decay, we have compared MC efficiencies using the two different models 
mentioned above: 
the one obtained by CLEO \cite{cleo2} (used in the final result for the 
branching ratio) 
and the same model but without interference: only the branching fractions 
of the intermediate resonances have been taken into account. Both models 
consider three intermediate resonances 
($\rho^+$, $\rho^-$, $\rho^0$ and a nonresonant 
contribution). \\ 

 For the $D^0 \to K\pi\pi^0$ mode results were 
obtained by CLEO \cite{kpp2} in the framework of a 7-resonance model   
($\rho^+$, ${K^*}^-$, ${\bar{K^0}}^*$, $K_0(1430)^-$, 
${\bar K}_0(1430)^0$,   
$\rho(1700)^+$, ${K^*}^-(1680)$ and a nonresonant contribution) and the 
3-resonance model (${K^*}^-\pi^+$, 
${\bar{K^0}}^* \pi^0$, $\rho^+\pi^-$, nonresonant $K^-\pi^+\pi^0$). 
The difference in the resulting efficiency using two different models 
is treated as a corresponding systematic error. The resulting systematic 
error due to the decay model uncertainty 
is 1.8\%. \\



As mentioned above, we use the generic MC shape in the 
$M(\pi^+\pi^-\pi^0)$ 
signal region as the default background description 
(when calculating the central value of 
${\cal B}$($D^0 \to \pi^+\pi^-\pi^0$)). To estimate the systematic error 
related to the description of background, we use an alternative function 
for the background fit. To do so, we extract true $K\pi\pi^0$ events 
from generic MC reconstructed as 
$\pi^+\pi^-\pi^0$. The shape obtained (see Fig. 5, right) is fitted with a 
$2^{nd}$ order polynomial 
multiplied by an error function and a $1^{st}$ order polynomial. 
We use this parameterization for the misidentification 
background, while another $1^{st}$ order polynomial is added to fit the 
other charm-- (other than misidentification), light-- and 
$b$--quark linear contributions. The difference between this and the 
default fit (0.4\%) is added in quadrature to the systematic error. \\ 



 To determine the signal fit uncertainty, a different 
$D^0 \to \pi^+\pi^-\pi^0$ signal parameterization, composed of two
bifurcated Gaussians and one regular Gaussian is used. 
For the $D^0 \to$ $K\pi\pi^0$ decay channel we use a bifurcated hyperbolic 
Gaussian and a regular Gaussian. The relative differences to the default fit 
are found to be 0.3\% and 0.5\% for the $\pi^+\pi^-\pi^0$ and 
$K^-\pi^+\pi^0$ mode, respectively, 
and are added to the total systematic uncertainty. \\ 


Finally, we varied the selection criteria and
estimate the systematic error corresponding to the possible inadequate
background description. The particle identification selection was
changed to $\mathcal{R}_{\rm PID}(\pi)<0.2$, 
$\mathcal{R}_{\rm PID}(K)>0.8$, 
and the resulting systematic uncertainty is found to be
negligible. Change of the $p_{\rm cms}$($D^*$) requirement (3.0 GeV/c) 
results in a 0.4\% error. A significant uncertainty is due to the Ks veto;
after excluding the region 
0.195 GeV$^2$/c$^4 \ < \ M^2(\pi^+\pi^-) \ <$  0.305 GeV$^2$/c$^4$, 
the resulting $\pi^+\pi^-\pi^0$ yield changes by 1.6\%. \\ 

%


The individual sources of the systematic error are listed in Table \ref{t2}. 
The total uncertainty is obtained by adding all contributions in quadrature. 

\begin{table}[ht!] 
\caption{Systematic errors} 
\vspace{0.5cm} 
\label{t2} 
\begin{tabular}{|l|l|} 
 \hline 
Source & Error, \% \\
\hline
 MC statistics                   &  0.8 \\ 
 PID efficiency of $K$/$\pi$       &  1.6 \\ 
 Decay model                     &  1.8 \\ 
 Fit (background $\&$ signal)    &  0.7 \\ 
 $p_{cms}$($D^*$) cut            &  0.4 \\ 
 $K_S^0$ veto                         &  1.6 \\ \hline 
 Total                           &  3.1 \\ \hline 
\end{tabular} \\ 
\end{table} 

\section{Results} 

Summarizing the discussion above, we obtain 
the following ratio of the branching fractions: 

$$\frac{{\cal B}(D^0 \to \pi^+\pi^-\pi^0)} 
{{\cal B}(D^0 \to K^-
\pi^+\pi^0)} \ = \ 
\frac{{\rm N}_{\pi^+\pi^-\pi^0}}{{\rm N}_{K^-\pi^+\pi^0}} \cdot 
\frac{\varepsilon_{K\pi\pi^0}}{\varepsilon_{\pi^+\pi^-\pi^0}} 
\cdot \frac{r_K}{r_{\pi}} 
 \ = \  0.0971 \pm 0.0009_{\rm stat} \pm 0.0030_{\rm syst}. $$ \\ 



 We multiply the obtained value by the 2006 world avearge of 
${\cal B}(D^0 \to K^-\pi^+\pi^0) \ = \ (14.1 \pm 0.5) \%$ \cite{pdg06} 
to calculate the absolute branching fraction for $D^0 \to \pi^+\pi^-\pi^0$ 
decay (see Table \ref{t4}). 
In a recent study by CLEO~\cite{cleo3} the relative branching ratio 
${\cal B}(D^0 \to \pi^+\pi^-\pi^0)$/${\cal B}(D^0\to K\pi)$ 
is measured to be 0.344 $\pm$ 0.013. 
The comparison of the corresponding values for the absolute branching ratio 
${\cal B}$($D^0 \to \pi^+\pi^-\pi^0$) shows that both results are in good agreement 
within the errors (see Table \ref{t4}). \\ 

\begin{table}[ht!] 
\caption{Values of the absolute branching fraction ${\cal B}$($D^0 \to \pi^+\pi^-\pi^0$) 
by Belle and CLEO-c.} 
\vspace{0.5cm} 
\label{t4} 
\begin{tabular}{|l|c|c|} 
\hline 
         & N$_{\rm ev.}$ & ${\cal B}$($D^0 \to \pi^+\pi^-\pi^0$), 10$^{-3}$ \\ \hline 
Belle    & 22803 $\pm$ 203  &  13.69 $\pm$ 0.13$_{\rm stat}$  
	 $\pm$ 0.42$_{\rm syst}$ $\pm$ 0.49$_{\rm norm}$  \\ 
CLEO-c   & 10834 $\pm$ 164  &  13.2 $\pm$ 0.2$_{\rm{stat}}$ 
	 $\pm$ 0.5$_{\rm{syst}}$ $\pm$ 0.2$_{\rm{norm}}$ $\pm$ 0.1$_{\rm CP corr.}$ \\ 
	 \hline 
\end{tabular} \\ 
\end{table} 


 Finally, we can compare our measurement of the ratio with the 
ratio obtained from the latest world average values of 
$\frac{{\cal B}(D^0 \to \pi^+\pi^-\pi^0)}{{\cal B}(D^0\to K^-\pi^+)}$ 
and 
${\cal B}$($D^0 \to \pi^-\pi^+\pi^0$) \cite{pdg06}. 
Our result (9.71 $\pm$ 0.31)\% 
is consistent with the world average one (9.29 $\pm$ 0.54)\% and has 
better accuracy. 
By choosing the normalization mode $D^0 \to K^-\pi^+\pi^0$ we avoid many 
sources of systematic uncertainty including the $\pi^0$ detection efficiency 
and uncertainty in the tracking efficiency. The PID efficiency partially 
cancels out. \\ 

\section{Summary} 

Using 357 fb$^{-1}$ of data collected with the Belle detector,
the first direct measurement of the relative branching fraction
${\cal B}$($D^0 \to \pi^+\pi^-\pi^0$)/
${\cal B}$($D^0 \to K^-\pi^+\pi^0$) has been performed. 
Our preliminary result $0.0971 \pm 0.0031$ 
is compatible with the world average $0.0929 \pm 0.0054$ and is 
more precise. The Belle result differs by $\sim2.8 \ \sigma$ from 
the value recently obtained by BaBar \cite{babar}: 
$0.1059 \pm 0.0006 \pm 0.0013$. 
The corresponding value of the  absolute branching fraction 
${\cal B}$($D^0 \to \pi^+\pi^-\pi^0$) is 
(13.69 $\pm$ 0.66)$\times 10^{-3}$. \\ 


\section*{Acknowledgments}

We thank the KEKB group for the excellent operation of the
accelerator, the KEK cryogenics group for the efficient
operation of the solenoid, and the KEK computer group and
the National Institute of Informatics for valuable computing
and Super-SINET network support. We acknowledge support from
the Ministry of Education, Culture, Sports, Science, and
Technology of Japan and the Japan Society for the Promotion
of Science; the Australian Research Council and the
Australian Department of Education, Science and Training;
the National Science Foundation of China and the Knowledge
Innovation Program of the Chinese Academy of Sciences under
contract No.~10575109 and IHEP-U-503; the Department of
Science and Technology of India;
the BK21 program of the Ministry of Education of Korea,
the CHEP SRC program and Basic Research program
(grant No.~R01-2005-000-10089-0) of the Korea Science and
Engineering Foundation, and the Pure Basic Research Group
program of the Korea Research Foundation;
the Polish State Committee for Scientific Research;
the Ministry of Science and Technology of the Russian
Federation; the Slovenian Research Agency;  the Swiss
National Science Foundation; the National Science Council
and the Ministry of Education of Taiwan; and the U.S.\
Department of Energy.

%
%
%


\begin{thebibliography}{99}
\bibitem{par} {The same function
 for MC is used to fit the data signak peak. The means and the relative normaliza
 of the constituting Gaussians are fixed to MC, $\sigma$'s and the overall normal
 are left floating.}
\bibitem{hyp} {Hyperbolic Gaussian:  
 $f = \frac{2 N}{2 \sqrt{\pi}\sigma}
 {\alpha |\frac{x-x_0}{\sigma}|}/(y \sqrt{y-\alpha}) \ \exp(-(y-\alpha))$,  
 $y \ = \ \sqrt{\alpha^2 + \alpha (\frac{x-x_0}{\sigma})^2}$,  
 where the parameters of the distribution are $N$ (normalization),  
 $x_0$ (mean, or peak position), $\sigma$ (width) and $\alpha$.  
 $f((\frac{x-x_0}{\sigma})^2 \ll \alpha) \ \approx \ 
 \frac{N}{\sqrt{2 \pi}\sigma} \exp(-(\frac{x-x_0}{\sigma})^2)$,   
 $ \ f((\frac{x-x_0}{\sigma})^2 \gg \alpha) \ \approx \ 
 \frac{N}{\sqrt{\pi\sigma}} \frac{1}{|x-x_0|} \exp(-\sqrt{\alpha}|\frac{x-x_0}{\sigma}|)$.}
\bibitem{bif} {Bifurcated Gaussian: 
 $f = \frac{2 N}{\sqrt{2 \pi}(\sigma_1 + \sigma_2)} \exp(-(\frac{x-x_0}{\sigma})^2)$,  
 $\sigma(x < x_0) = \sigma_1$, $\sigma(x > x_0) = \sigma_2$, the parameters of the distribution
 are $N$ (normalization), $x_0$ (peak position), $\sigma_1$ and $\sigma_2$ (widths).}
\end{thebibliography}
\end{document}